# Instant tissue field and magnetic susceptibility mapping from MRI raw phase using Laplacian enhanced deep neural networks

Yang Gao[1], Zhuang Xiong[1], Amir Fazlollahi[2], Peter J Nestor[2], Viktor Vegh[3,4], Fatima Nasrallah[2], Craig Winter[5,6,7], G. Bruce Pike[8], Stuart Crozier[1], Feng Liu[1], Hongfu Sun[1,*]

[1]School of Information Technology and Electrical Engineering, University of Queensland, Brisbane, Australia

[2]Queensland Brain Institute, University of Queensland, Brisbane, Australia

[3]Centre for Advanced Imaging, University of Queensland, Brisbane, Australia

[4]ARC Training Centre for Innovation in Biomedical Imaging Technology, Brisbane, Australia

[5]Kenneth G Jamieson Department of Neurosurgery, Royal Brisbane and Women's Hospital, Brisbane, Australia

[6]Centre for Clinical Research, University of Queensland, Brisbane, Australia

[7]School of Mechanical, Medical and Process Engineering, Queensland University of Technology, Brisbane, Australia

[8]Departments of Radiology and Clinical Neurosciences, Hotchkiss Brain Institute, University of Calgary, Calgary, Canada

[*]**Correspondence**: Hongfu Sun

**Address**: Room 540, General Purpose South (Building 78), University of Queensland, St Lucia QLD 4072, Australia

**Email**: hongfu.sun@uq.edu.au

**Abbreviations:**

**iQSM**, instant quantitative susceptibility mapping; **iQFM**, instant quantitative field mapping; **LoT**, Laplacian-of-Trigonometric-functions; **ME-GRE**, multi-echo gradient-echo; **FOV**, field of view; **ROI**, region-of-interest; **DGM**, deep grey matter; **FWM**, frontal white matter; **LBV**, Laplacian Boundary Value; **PDF**, Projection onto Dipole Fields; **RESHARP**, regularization enabled sophisticated harmonic artifact reduction for phase data; **TKD**, truncated dipole inversion; **TGV-QSM**, total-generalized-variation QSM; **CNN**, convolutional neural network; **BET**, brain extraction tool; **FSL**, FMRIB Software Library; **COSMOS**, Calculation Of Susceptibility through Multiple Orientation Sampling; **GP**, globus pallidus; **PU**, putamen; **CN**, caudate nucleus; **TH**, thalamus; **SN**, substantia nigra; **RN**, red nucleus; **PSNR**, peak signal to noise ratio; **SSIM**, structural similarity; **NRMSE**, normalized root-mean-squared error; **MS**, multiple sclerosis; **ICH**, intracranial hemorrhage; **ppb**, part-per-billion.

# ABSTRACT

Quantitative susceptibility mapping (QSM) is an MRI post-processing technique that produces spatially resolved magnetic susceptibility maps from phase data. However, the traditional QSM reconstruction pipeline involves multiple non-trivial steps, including phase unwrapping, background field removal, and dipole inversion. These intermediate steps not only increase the reconstruction time but accumulates errors. This study aims to overcome existing limitations by developing a Laplacian-of-Trigonometric-functions (LoT) enhanced deep neural network for near-instant quantitative field and susceptibility mapping (i.e., iQFM and iQSM) from raw MRI phase data. The proposed iQFM and iQSM methods were compared with established reconstruction pipelines on simulated and *in vivo* datasets. In addition, experiments on patients with intracranial hemorrhage and multiple sclerosis were also performed to test the generalization of the proposed neural networks. The proposed iQFM and iQSM methods in healthy subjects yielded comparable results to those involving the intermediate steps while dramatically improving reconstruction accuracies on intracranial hemorrhages with large susceptibilities. High susceptibility contrast between multiple sclerosis lesions and healthy tissue was also achieved using the proposed methods. Comparative studies indicated that the most significant contributor to iQFM and iQSM over conventional multi-step methods was the elimination of traditional Laplacian unwrapping. The reconstruction time on the order of minutes for traditional approaches was shortened to around 0.1 seconds using the trained iQFM and iQSM neural networks.

## 1. INTRODUCTION

Quantitative susceptibility mapping (QSM) is a post-processing technique involving the estimation of the body tissue susceptibility distribution from MR phase images (de Rochefort et al., 2008; Deistung et al., 2013; Langkammer et al., 2012; Liu et al., 2015; Wang & Liu, 2015). The technique has shown significant potential for studying various neurological disorders, including Parkinson's disease (Acosta-Cabronero et al., 2017; Langkammer et al., 2016), Alzheimer's disease (Acosta-Cabronero et al., 2013; Ayton et al., 2017), multiple sclerosis (Elkady et al., 2018; Elkady et al., 2019; Langkammer et al., 2013; Wisnieff et al., 2015), intracranial hemorrhage (Sun et al., 2018a; Wang et al., 2013), alcohol use disorders (Juhas et al., 2017), and healthy aging (Bilgic et al., 2012; Chen et al., 2021b; Treit et al., 2021).

Different algorithms have been proposed for each QSM reconstruction step, including best-path (Abdul-Rahman et al., 2009) and Laplacian-based (Li et al., 2014) methods for phase unwrapping; PDF (Liu et al., 2011), SHARP (Schweser et al., 2011), RESHARP (Sun & Wilman, 2014), and LBV (Zhou et al., 2014) for background field removal; as well as TKD (Shmueli et al., 2009), iLSQR (Li et al., 2015) and MEDI (Liu et al., 2012) for dipole inversion. In addition, single-step TGV-QSM (Chatnuntawech et al., 2017; Langkammer et al., 2015) and total inversion methods (Sun et al., 2018b) have also been proposed to eliminate some of the intermediate post-processing steps. The continued development of the QSM pipeline is evidence of methodological shortcomings. Additionally, the highly involved post-processing steps make QSM methods computationally intensive and in turn slow whilst requiring manual input for parameter tuning. Moreover, conventional methods can lead to overwhelming artifacts in image regions of large susceptibility sources, particularly those around intracranial hemorrhage (Sun et al., 2016), where MR signal decays highly rapidly and phase wraps become dense and severe.

The final post-processing step of phase images within the traditional QSM pipeline is the solution to the field-to-susceptibility inverse problem, often referred to as dipole inversion. An increasing number of deep learning methods have been applied for QSM dipole inversion. QSMnet (Yoon et al., 2018) trained an Unet (Ronneberger et al., 2015) with single-orientation acquired local field and multiple-orientation reconstructed susceptibility (i.e., COSMOS (Chen et al., 2021a; Liu et al., 2009)) as inputs and labels. The method was later extended to QSMnet+ (Jung et al., 2020) to improve reconstruction using a data augmentation technique. QSMGAN (Chen et al., 2020), VaNDI (Polak et al., 2020), LPCNN (Lai et al., 2020), and MoDL-QSM (Feng et al., 2021) further improved this training scheme through the implementation of more advanced network structures. DeepQSM (Bollmann et al., 2019b) and xQSM (Gao et al., 2021) trained deep neural networks using the susceptibility-to-field forward model so that the training inputs and labels satisfy the underlying dipole convolution equation. AutoQSM (Wei et al., 2019) trained Unet to take total field maps as inputs and STAR-QSM (Wei et al., 2015) as labels, skipping the background field removal step. Apart from solving the ill-posed dipole inversion, deep learning methods for other QSM processing steps have been designed, including PHU-NET (Zhou et al., 2021) for phase unwrapping and SHARQnet (Bollmann et al., 2019a) for background field removal.

This work proposes a novel network architecture, which provides a mathematical description for the background field, typically removed prior to dipole inversion. The result is a single-step method that takes raw phase images of human brains as input and estimates spatially resolved maps of tissue field and magnetic susceptibility. The proposed methods were compared with multiple established reconstruction pipelines on simulated and *in vivo* human brain datasets. In addition, experiments on patients with intracranial hemorrhage and multiple sclerosis were also performed to test the generalizability of the method.

## 2. METHOD

### 2.1. Laplacian of trigonometric functions on phase

The relationship between magnetic susceptibility ($\chi$) and unwrapped phase ($\varphi_u$) is described through the Laplacian operator (Li et al., 2014; Salomir et al., 2003), leading to a second-order partial differential equation:

$$\frac{1}{3}\frac{\partial^2 \chi}{\partial x^2} + \frac{1}{3}\frac{\partial^2 \chi}{\partial y^2} - \frac{2}{3}\frac{\partial^2 \chi}{\partial z^2} = \frac{1}{2\pi\gamma B_0 TE}\nabla^2 \varphi_u, \tag{1}$$

where $\nabla^2$ (nabla-squared) is the Laplacian operator; $B_0$ is the main magnetic field; $TE$ is the echo time, and $\gamma$ is the gyromagnetic ratio for hydrogen.

The Laplacian of the unwrapped phase can be calculated from the raw wrapped phase (Schofield & Zhu, 2003):

$$\nabla^2 \varphi_u = \cos(\varphi_w) \cdot \nabla^2 \sin(\varphi_w) - \sin(\varphi_w) \cdot \nabla^2 \cos(\varphi_w), \tag{2}$$

where $\varphi_w$ is the wrapped phase. Equations (1) and (2) have formed the basis of integrated phase unwrapping and background phase removal (Li et al., 2014; Schweser et al., 2013) and single-step dipole inversion (Chatnuntawech et al., 2017; Langkammer et al., 2015).

## 2.2. LoT-Unet neural network

A novel deep neural network architecture, LoT-Unet, incorporating a Laplacian-of-Trigonometric-functions (LoT) layer followed by a 3D residual Unet, is shown in Fig. 1. We use the same basic network architecture for instant quantitative field mapping (iQFM) and quantitative susceptibility mapping (iQSM). The proposed LoT layer is designed to process the wrapped raw phase using Eq. (2). In the discrete form, it becomes:

$$LoT = \frac{cos(\varphi_w) \cdot (K * sin(\varphi_w)) - sin(\varphi_w) \cdot (K * cos(\varphi_w))}{B_0 \cdot TE}, \qquad (3)$$

where $K \in \mathbb{R}^{3 \times 3 \times 3}$ is a 27-point stencil discrete Laplace kernel (O'Reilly & Beck, 2006) replacing the continuous $\nabla^2$ operator in Eq. (2), and it takes a fixed form in the network; $LoT$ denotes the output from the LoT layer, which is the Laplacian of the unwrapped phase normalized by the main field $B_0$ and echo time TE. The asterisk ($*$) in Eq. (3) denotes the convolution operation, while dot ($\cdot$) represents element-wise multiplication. The 27-point discrete Laplacian operator:

$$\text{first or third plane: } 1/13 \begin{bmatrix} 1 & 3/2 & 1 \\ 3/2 & 3 & 3/2 \\ 1 & 3/2 & 1 \end{bmatrix}, \text{ second plane: } 1/13 \begin{bmatrix} 3/2 & 3 & 3/2 \\ 3 & -44 & 3 \\ 3/2 & 3 & 3/2 \end{bmatrix},$$

can be incorporated as a tailored convolutional layer of the whole neural network and accelerated by GPU. The 3D Unet, widely used in previous deep learning-based QSM studies, is applied after the LoT layer to recover the tissue field or susceptibility, depending on the training labels. It consists of 18 convolutional layers (kernel size: 3×3×3), 4 max-pooling layers (kernel size: 2×2×2), 4 transposed convolutional layers (kernel size: 2×2×2), 22 batch-normalization layers, 22 rectified linear units (ReLUs), 4 concatenation layers and 1 final convolutional layer (kernel size: 1×1×1). In addition, a skip connection was added between the input and the output, forming a residual block (Kim et al., 2016), which helps the training process converge faster. Furthermore, this skip connection can mitigate the vanishing gradient problem during training (He et al., 2016) and provide a noticeable performance enhancement (Jin et al., 2017).

The LoT layer is designed to carry out the Laplacian operations following Eq. (2), thus the LoT convolutional kernels ($K$ in Eq. (3)) are precisely set to the 27-point stencil discrete Laplacian formation. However, since the neural network approach is data-driven, we can also set these fixed Laplacian kernel weights as "learnable" during network training. This point is detailed in the Discussion section.

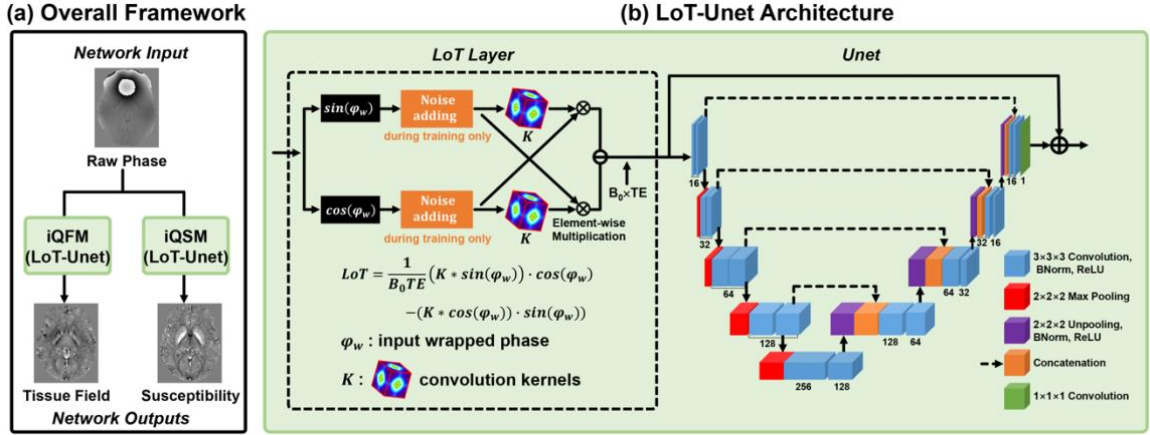

**Figure 1.** Overview of iQFM and iQSM framework using the proposed LoT-Unet architecture, composed of a tailored LoT layer and a 3D residual Unet. Two independent noise adding blocks (orange boxes) were added for the real and imaginary components of the input during network training. These noise-adding blocks were removed from the successfully trained network for QSM reconstruction tasks during network deployment.

For multi-echo GRE data, iQFM and iQSM networks are applied on individual echoes, followed by a voxel-wised magnitude-weighted least-square fitting with the echo times (as described in Eq. (4) below) to obtain the final local field and susceptibility maps, as demonstrated in Supp. Fig. 1.

$$x = (E^T M E)^{-1} E^T M Y, \qquad (4)$$

where $x$ is the echo-fitted results; $E = [TE_1, TE_2 \cdots, TE_N]^T$ represents the echo time vector from $N$ echoes; $M$ is a diagonal matrix of magnitude intensities from all echoes; $Y = [y_1, y_2 \cdots, y_N]^T$ is TE-scaled local field or susceptibility.

### 2.3. Training datasets

Institutional ethics board approval was obtained, and all subjects gave informed written consent. Training datasets were acquired from 96 healthy subjects with a 3D multi-echo GRE sequence at 3T (GE Discovery 750) with scan parameters: 8 unipolar echoes, first TE = 3 ms, echo spacing = 3.3 ms, TR = 29.8 ms, FOV = 144×196×128 mm³, isotopic voxel size = 1 mm³. An established QSM reconstruction pipeline, i.e., Laplacian unwrapping followed by RESHARP background removal (with 3-voxel brain edge erosion applied on the brain masks generated with BET (Smith, 2002)) and then total variation dipole inversion (Bilgic et al., 2012), was applied to generate susceptibility maps as the training labels. As illustrated in Fig. 2, a forward dipole convolution was conducted on the susceptibility map (i.e., *in vivo*) to simulate the local field. Then, the corresponding background field maps, extracted from the total field maps (i.e., *in vivo*) using PDF background field removal, were added to the simulated local field images to obtain the simulated total field datasets. Finally, wrapped phase maps were produced by linearly scaling with random echo times and fixed field strength of 3T. The random TE setting was made normally distributed with a mean of 20 ms and a standard deviation of 10 ms.

Note that the forward dipole convolution was carried out on small susceptibility patches (size: 64×64×64) instead of the whole brain volumes (size: 144×196×128). A previous study (Zhu et al., 2021) showed that this led to improved performance compared to computing dipole convolution on the whole brain volumes. In addition, dipole convolutions on small QSM patches also made it more flexible to include any number of pathological sources into the training datasets (detailed below). The cropping was carried out by sliding a 64×64×64 window with a stride of 16×26×21 to traverse each full-size QSM volume, resulting in 13,824 patches from all 96 patients. This approach retains more data features from the full-sized QSM volumes and eliminates potential patch similarities and repetitions than using random cropping.

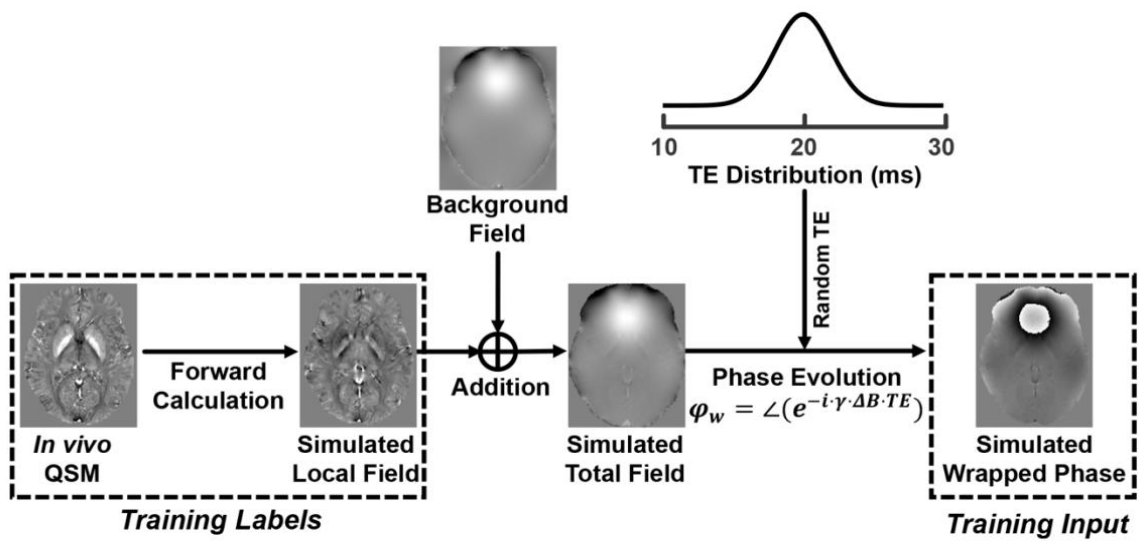

**Figure 2**. Training data preparation pipeline, including a forward field calculation to simulate local field and a phase evolution process to obtain wrapped phase.

Another set of susceptibility patches were simulated, with each patch including either a hemorrhage or a calcification lesion. These hemorrhage and calcification sources were generated as follows. First, images of matrix size 16×16×16, containing basic geometric shapes, were constructed. Each image has 5-10 spheres, 5-10 rectangles, and 5-10 cubes overlayed, with each geometric shape size randomly set to between 10% and 40% of the image size. Next, these images were binarized and randomly resized to matrix sizes from 12×12×12 to 24×24×24. Then, each binary image was multiplied by a constant to represent either a hemorrhage or a calcification susceptibility source, with equal probability for either type. This constant susceptibility was randomly assigned from uniform distributions with ranges of [0.4, 1.2] and [-0.3, -0.1] for hemorrhage (Sun et al., 2018a) and calcification (Barbosa et al., 2015), respectively. Supplemental Figure 2 illustrates two examples of these training patches with simulated pathological sources. Finally, these simulated "pathological" susceptibility sources were superpositioned onto the "healthy" susceptibility patches at random positions to obtain the other set of training data.

## 2.4. Network training

For each epoch, a total of 13,824 patches (60% healthy patches and 40% pathological patches from different cropping locations) were randomly selected from the two datasets (i.e., healthy and pathological) for training. The iQSM network was trained with simulated wrapped phase as inputs and susceptibility as labels, while iQFM was trained with wrapped phase as inputs and local field as labels. The two networks were trained individually, and the trained network models were also applied independently for local field and susceptibility reconstructions.

The networks were implemented using Pytorch 1.8; the trained networks and the source codes are available at https://github.com/sunhongfu/deepMRI/tree/master/iQSM. Parameters in Unet were initialized with values pulled from a normal distribution with a mean of zero and a standard deviation of 0.01. The networks were trained using the Adam optimizer for 100 epochs with a mini-batch size of 32. The mean-squared-error loss function was used to train the networks. The learning rate was set to $10^{-3}$, $10^{-4}$, and $10^{-5}$ for the first 50, 50-80, and the last 20 epochs. In addition, two independent noise adding blocks detailed in a recent work (Gao et al., 2021) were applied to the real and imaginary parts of the complex phase data to improve the noise robustness of the network, as shown in the orange boxes in Fig. 1. The proportion of noisy patches during each epoch training was set to 20%, empirically. Note that the two noise adding blocks were only added during network training and were removed when reconstructing full-size QSM images during inference. It took about 20 hours to train either iQFM or iQSM on 2 Nvidia Tesla V100 GPUs. The reconstruction implementation of iQFM and iQSM for the evaluation data were conducted on the full-size images without cropping.

### 2.5. Evaluation datasets

The proposed iQFM and iQSM methods were evaluated on (1) two simulated brains, each with and without hemorrhage and calcification, (2) four *in vivo* healthy subjects, (3) three *in vivo* intracranial hemorrhage (ICH) patients, and (4) two *in vivo* multiple sclerosis (MS) patients, and (5) another *in vivo* healthy subject with multiple head orientations. The acquisitions, reconstructions, and comparisons for all evaluation datasets are detailed below.

(1) Raw phase data of two healthy brains were simulated from two COSMOS images using the pipeline shown in Fig. 2 to investigate the effects of the proposed LoT layer. Furthermore, to compare different methods on high susceptibility sources, a simulated hemorrhage (1 ppm uniform) and a calcification (-0.2 ppm uniform) were superimposed onto the healthy brains to simulate pathological brain phase data.

(2) Four healthy volunteers were scanned with the multi-echo GRE sequence of 1 mm isotropic resolution. Two of these subjects were scanned at 7T (4 unipolar echoes, first TE = 5 ms, echo spacing = 3 ms, TR = 15.5 ms, FOV = 192×256×144 mm$^3$), while the other two subjects were scanned at 3T with the same sequence parameters as already described for the training datasets. For the 7T scans acquired with a 32-channel head coil, the POEM method (Sun et al., 2020) was used to combine the raw multi-channel phase, and the channel-combined phase was input into the proposed neural networks.

(3) Multi-echo GRE acquisitions with five different head orientations (i.e., neutral head position, left 23°, right 17°, flexion 18° and extension 21° tilt angles from the neutral head position) were obtained from one healthy subject at 3T. Local field maps for each head orientation were produced through the two-step pipelines (Laplacian unwrapping + different background field removal methods) and the proposed iQFM method. The four local field maps of oblique head orientations were registered to the neutral orientation, and closed-form COSMOS maps were obtained according to the original paper (Liu et al., 2009). COSMOS results were compared on two brain masks (i.e., 1-voxel and 3-voxel brain edge erosions). The brain edge erosions were performed on the raw phase for iQFM while on the unwrapped phase before the background field removal for RESHARP, LBV, PDF, and SHARQnet.

(4) Three intracranial hemorrhage patients were scanned using 3T MRI at different medical centres. Subject #1: 7 unipolar echoes, first TE = 5.84 ms, echo spacing = 4.79 ms, TR = 40 ms, FOV = 180×224×144 mm$^3$, isotropic voxel size = 1 mm$^3$; Subject #2: 9 unipolar echoes, first TE = 5.80 ms, echo spacing = 4.80 ms, TR = 50 ms, FOV = 216×224×120 mm$^3$, isotropic voxel size: 1 mm$^3$; Subject #3: 4 unipolar echoes, first TE = 6.76 ms, echo spacing = 6.24 ms, TR = 35 ms, FOV = 256×256×128 mm$^3$, isotropic voxel size = 1 mm$^3$.

(5) Raw phase data from two multiple sclerosis patients were acquired with the same imaging parameters as the training datasets (detailed in Section 2.3 above) to visualize and delineate lesions in QSM results.

## 2.6. Performance evaluation and region-of-interest (ROI) measurements

The proposed iQFM and iQSM methods were compared with several established single-step and multi-step methods for tissue field and susceptibility reconstructions, including TGV-QSM (Langkammer et al., 2015), RESHARP (Sun & Wilman, 2014), LBV (Zhou et al., 2014), PDF (Liu et al., 2011), SHARQnet (Bollmann et al., 2019a), iLSQR (Li et al., 2015), MEDI (Liu et al., 2012), QSMnet$^+$ (Jung et al., 2020), and autoQSM (Wei et al., 2019). All multi-step methods started from Laplacian-based (Li et al., 2014) unwrapping results in this study. Different regularization parameters were tested, and the ones led to the best visually-appealing QSM results were chosen, with $(\alpha_1, \alpha_0) = (0.003, 0.001)$ for TGV-QSM and $\lambda = 2000$ for MEDI. The deep learning-based SHARQnet, QSMNet$^+$, and autoQSM were downloaded from the original paper and trained by their authors. All deep learning-based reconstructions were performed using one Tesla V100 GPU (32 GB), while the reconstructions of all traditional methods were accomplished on one Intel(R) Core(TM) i5-10600 3.30GHz CPU. The iQFM and iQSM were implemented using Pytorch 1.8, while QSMnet$^+$, SHARQnet, and autoQSM were implemented with Tensorflow 1.14 as reported in the original papers.

For quantitative assessment of different methods, standard numerical metrics, including peak signal-to-noise ratio (PSNR), structural similarity index measure (SSIM), and normalized root-mean-squared error (NRMSE), were computed. In addition, susceptibilities of deep grey matter (globus pallidus (GP), putamen (PU), caudate (CN), substantia nigra (SN), and red nucleus

(RN)), as well as the lesions (hemorrhage and calcification), were measured with ROIs drawn manually using ImageJ software (National Institutes of Health, Bethesda, MD). Moreover, bar graphs and line profiles on multiple hemorrhage lesions were measured to compare the performances of different methods on the *in vivo* hemorrhage patients. Finally, the delineations of white matter lesions in the MS patients were also compared.

## 3. RESULTS

### 3.1. Simulation results

Figure 3 compares the reconstruction results of various local field and susceptibility mapping methods on one simulated subject with one added hemorrhage and one calcification. The upper two rows show the results of local field maps and their errors, while the bottom two rows display the results and errors of susceptibility maps. As evident in Fig. 3 error maps, iQFM and iQSM are the only two methods with no substantial artifacts (red arrows) or susceptibility underestimations of the hemorrhage source. All background field removal methods (i.e., RESHARP, LBV, PDF, SHARQnet) started from the phase-unwrapped total field maps. This study used a Laplacian-based method (Li et al., 2014) for phase unwrapping instead of a path-based method (Abdul-Rahman et al., 2009) to reduce phase errors from the hemorrhage. On the contrary, iQFM and iQSM reconstructed local field and susceptibility maps from the raw wrapped phase instantly without the phase unwrapping and background field removal steps. The conventional Laplacian unwrapping followed by RESHARP, LBV, and PDF methods displayed residual background field near the sinus, while the deep learning-based SHARQnet suppressed the tissue field contrast, as can be deduced from the error map. A substantial degree of deep grey matter susceptibility underestimation was also observed in the TGV-QSM method.

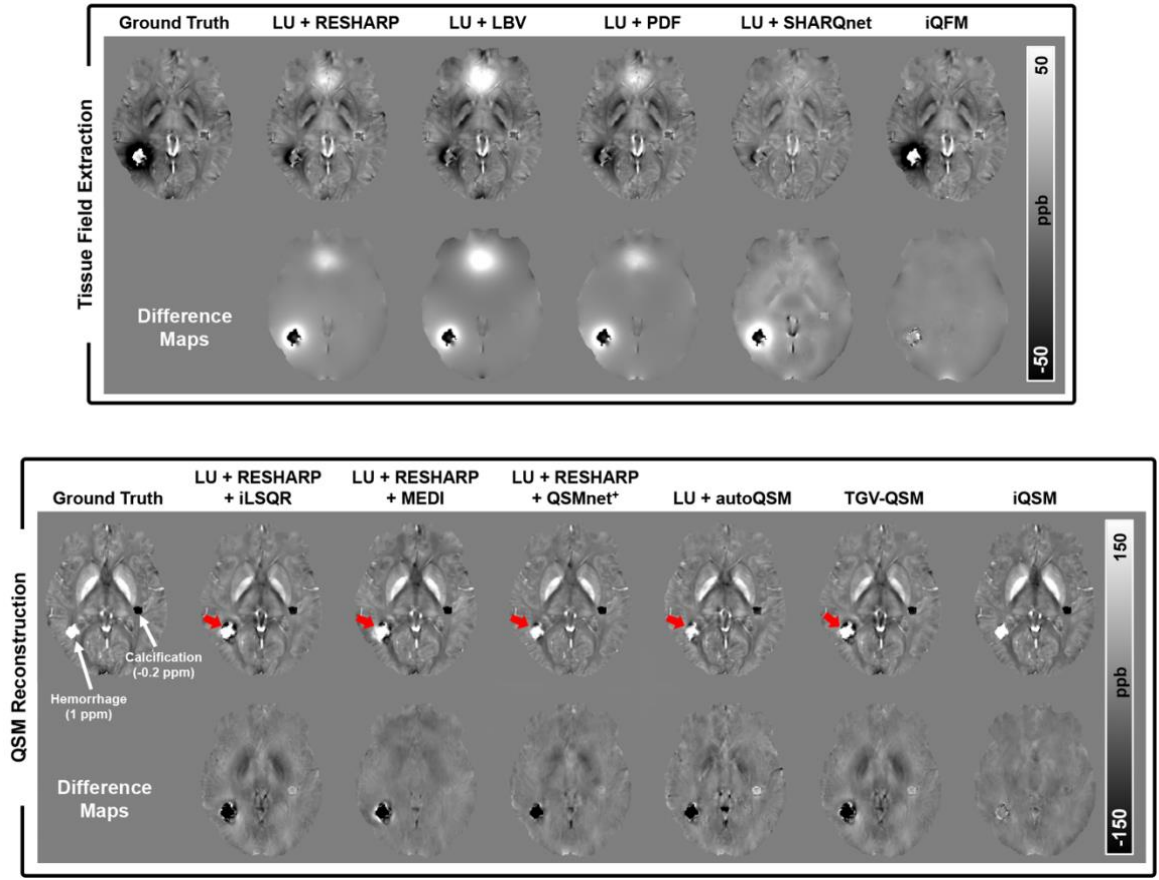

**Figure 3**. The comparison of quantitative field mapping (QFM) and QSM methods on one simulated pathological data. The upper two rows show the QFM results from different methods and their difference maps to the simulated ground truth, while the bottom two rows demonstrate the QSM results and the corresponding difference maps from the ground truth. White arrows locate the added hemorrhage and calcification, while red arrows point to artifacts near the hemorrhage source. LU denotes Laplacian unwrapping.

Numerical performance metrics (i.e., PSNR, SSIM, and NRMSE) of the different methods are reported for both healthy and pathological simulations, as reported in Table 1 (for local field mapping) and Table 2 (for QSM). For most cases, the proposed iQFM and iQSM led to the best PSNR and NRMSE and comparable SSIM with Laplacian unwrapping multi-step methods. More importantly, the NRMSEs of iQFM and iQSM are 2-4 times lower than other methods in the pathological brain with hemorrhage susceptibility assigned to 1ppm. In addition, the tissue field reconstruction time of this data (image size: 256×256×128) for iQFM is only 130 ms (on GPU), compared to 1 s for SHARQnet (on GPU) and 8 s for LBV (on CPU), 2 mins for PDF (on CPU), and 4 mins for RESHARP (on CPU). The susceptibility reconstruction time of iQSM is also 130 ms (on GPU), compared to 20 s for autoQSM (on GPU), 8 mins for TGV-QSM (on CPU), and 4 mins (Laplacian unwrapping + RESHARP background field removal on CPU) plus 12 s for QSMnet+ reconstruction (on GPU), 40 s for iLSQR reconstruction (on CPU), 120 s for MEDI reconstruction (on CPU).

To isolate errors originating from each reconstruction step in the conventional QSM pipeline, we performed a series of experiments with results reported in Fig. 4. It is shown that the dipole inversion step alone (i.e., iLSQR) led to an 11% underestimation of hemorrhage susceptibility. Background field removal followed by dipole inversion (i.e., RESHARP + iLSQR) introduced 17% hemorrhage susceptibility underestimation. Lastly, about 75% susceptibility underestimation was observed on the hemorrhage source when QSM was reconstructed from the raw phase through Laplacian phase unwrapping, RESHARP background field removal and finally iLSQR dipole inversion. It is evident from the difference maps in the total field and local field results that the most substantial errors in the hemorrhage region were from the Laplacian phase unwrapping step, and these errors can be dramatically alleviated with the proposed iQFM and iQSM methods.

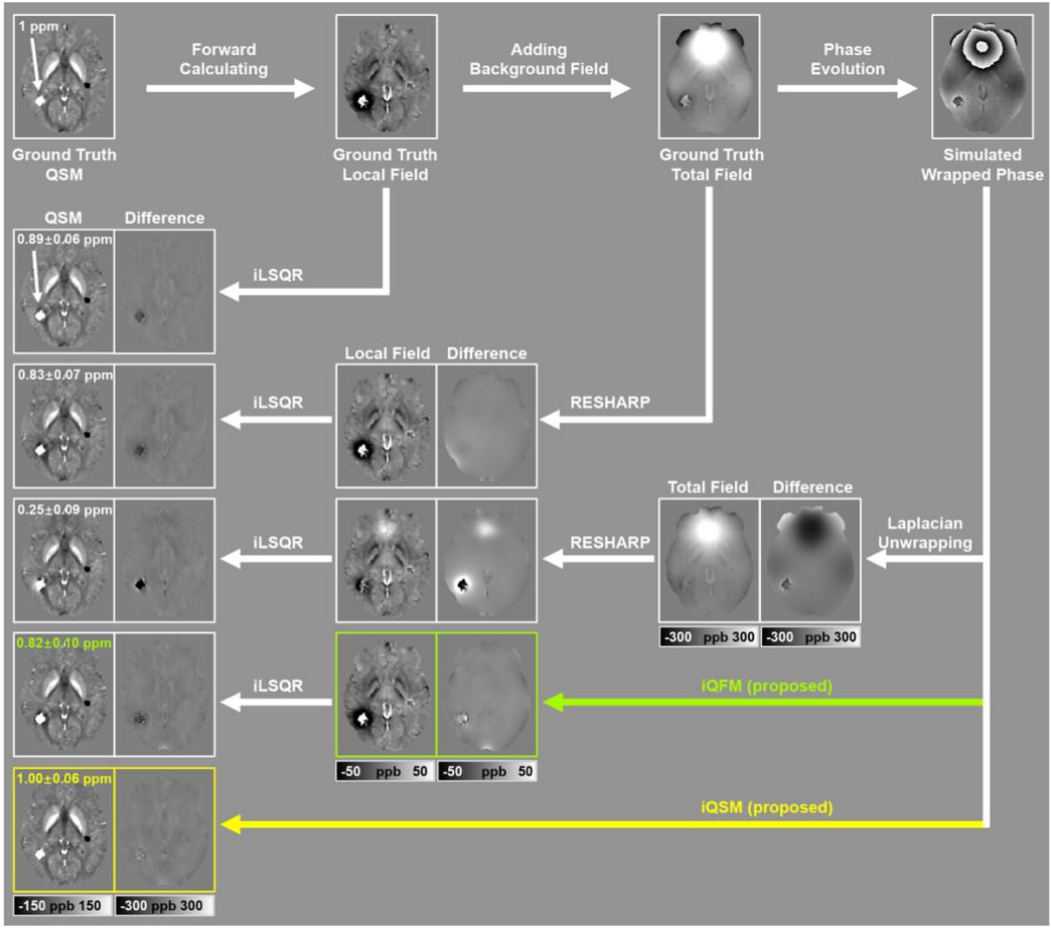

**Figure 4**. Illustration of the reconstruction errors from each step in a traditional QSM reconstruction pipeline (Laplacian unwrapping + RESHARP + iLSQR) and the proposed frameworks (iQFM + iLSQR and iQSM).

**Table 1**. Numerical performance metrics of different tissue field mapping methods on the healthy and pathological simulation data.

|  | PSNR | | SSIM | | NRMSE (%) | |
| --- | --- | --- | --- | --- | --- | --- |
|  | healthy | pathological | healthy | pathological | healthy | pathological |
| LU + RESHARP | 53.59 | 44.61 | 0.99 | 0.99 | 49.71 | 74.73 |
| LU + LBV | 47.93 | 43.55 | 0.99 | 0.98 | 82.23 | 83.05 |
| LU + PDF | 53.98 | 44.56 | 0.99 | 0.99 | 52.60 | 75.22 |
| LU + SHARQnet | 49.70 | 42.86 | 0.98 | 0.97 | 72.11 | 89.14 |
| iQFM | **56.32** | **52.47** | 0.99 | 0.99 | **36.52** | **37.31** |

The best results are highlighted in bold. LU denotes Laplacian unwrapping.

Table 2. Numerical performance metrics of different tissue susceptibility mapping methods on the healthy and pathological simulation data.

|  | PSNR | | SSIM | | NRMSE (%) | |
| --- | --- | --- | --- | --- | --- | --- |
|  | healthy | pathological | healthy | pathological | healthy | pathological |
| LU + RESHARP + iLSQR | 43.96 | 37.65 | **0.96** | 0.95 | 55.04 | 69.48 |
| LU + RESHARP + MEDI | 44.07 | 37.61 | 0.95 | 0.95 | 54.37 | 69.85 |
| LU + RESHARP + QSMnet[+] | 43.30 | 36.78 | 0.95 | 0.95 | 59.41 | 75.88 |
| LU + autoQSM | 43.27 | 36.44 | 0.94 | 0.94 | 66.85 | 79.85 |
| TGV-QSM | 34.88 | 31.18 | 0.87 | 0.87 | 160.19 | 146.14 |
| iQSM | **44.27** | **42.91** | 0.95 | 0.95 | **52.34** | **36.36** |

The best results are highlighted in bold. LU denotes Laplacian unwrapping.

Susceptibility measurements of the hemorrhage and calcification lesions and five deep grey matter regions from the two simulated subjects are compared in Fig. 5. It is seen that iQSM is the only method that accurately measured the hemorrhage susceptibility with only 0.4% deviation from the ground truth, while all other methods substantially underestimate hemorrhage susceptibility by about 80% of its actual value. Besides, iQSM also achieved the most accurate susceptibility reconstruction of the deep grey matter regions, especially for the globus pallidus, with only 2.5% mean error, compared to TGV-QSM with 40% susceptibility underestimation.

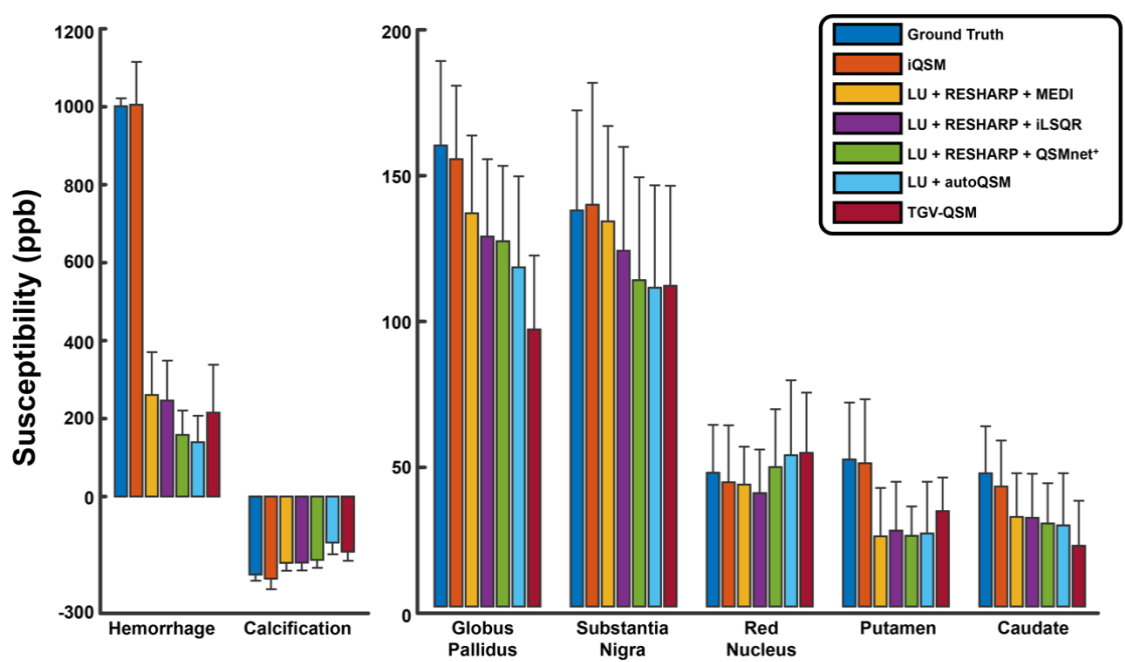

**Figure 5**. ROI analysis of different QSM reconstruction methods. The left panel shows the susceptibility measurements of the hemorrhage and calcification sources, while the right panel demonstrates the five deep grey matter susceptibilities. The iLSQR, MEDI, and QSMnet[+] were based on Laplacian unwrapping, followed by RESHARP background removal results, while autoQSM takes the Laplacian unwrapping results as inputs. Whiskers on the bars indicate the standard deviation of the susceptibility measurements from all voxels inside the ROI. LU denotes Laplacian unwrapping.

### 3.2. *In vivo* experiment results from healthy subjects

Figure 6 compares the proposed iQFM and iQSM with other single-step and multi-step conventional and deep learning methods on a healthy subject acquired *in vivo* at 7T with isotropic 1 mm$^3$ resolution. When comparing the local field maps, LBV exhibited severe artifacts (red arrows), SHARQnet showed substantial contrast loss (yellow arrows), while the iQFM resulted in a local field map free of artifact and contrast loss. Furthermore, iQSM, iLSQR, MEDI, and QSMnet[+] showed similar image contrasts as COSMOS results, while TGV-QSM and autoQSM showed considerable loss of susceptibility contrast as indicated by the yellow arrows. Finally, the deep grey matter susceptibility measurements from four healthy subjects are compared in Fig. 7. On average, the iQSM method led to 5.08% measurement deviation from COSMOS results in DGM regions, compared to 10.04%, 10.73%, 11.32%, 33.12%, and 48.78% for QSMnet[+], MEDI, iLSQR, autoQSM, and TGV-QSM, respectively, which is consistent with simulation results in Fig. 5.

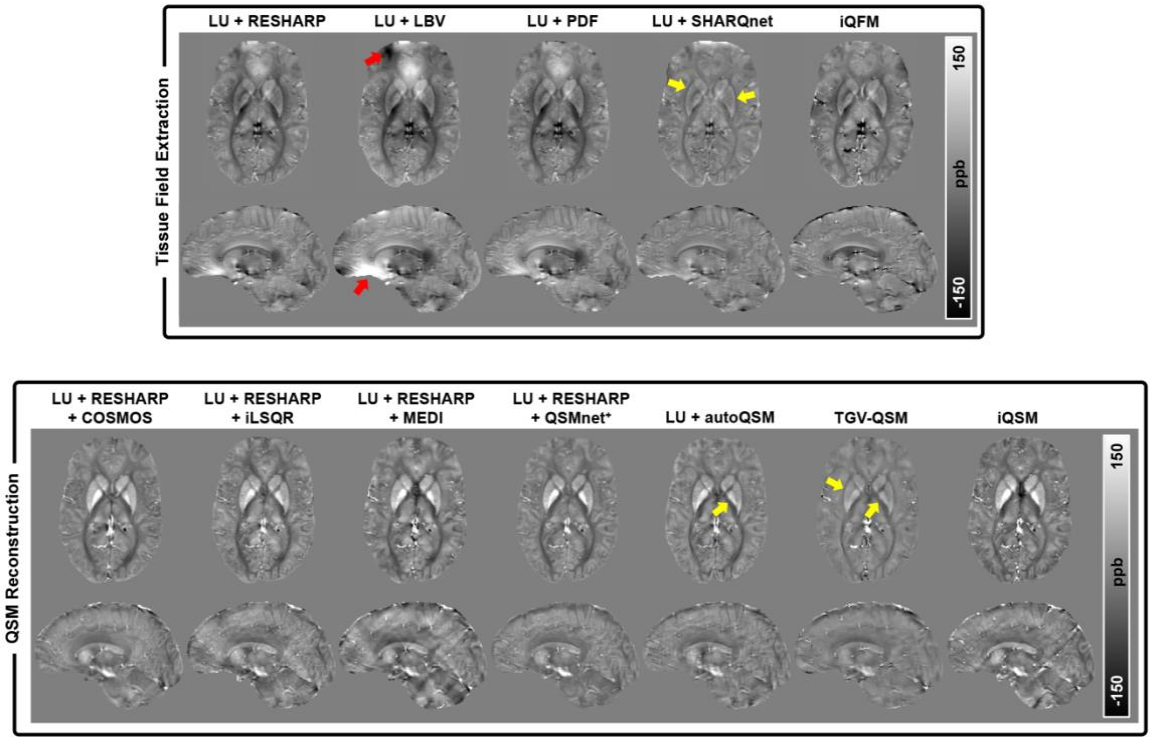

**Figure 6.** Comparison of different local field and susceptibility reconstruction methods on one *in vivo* subject at 7T. The upper panel demonstrates the tissue field results, while the bottom panel illustrates the QSM results. Red arrows point to the artifacts in the local field maps. Yellow arrows point to the contrast loss in some local fields and susceptibility results. LU denotes Laplacian unwrapping.

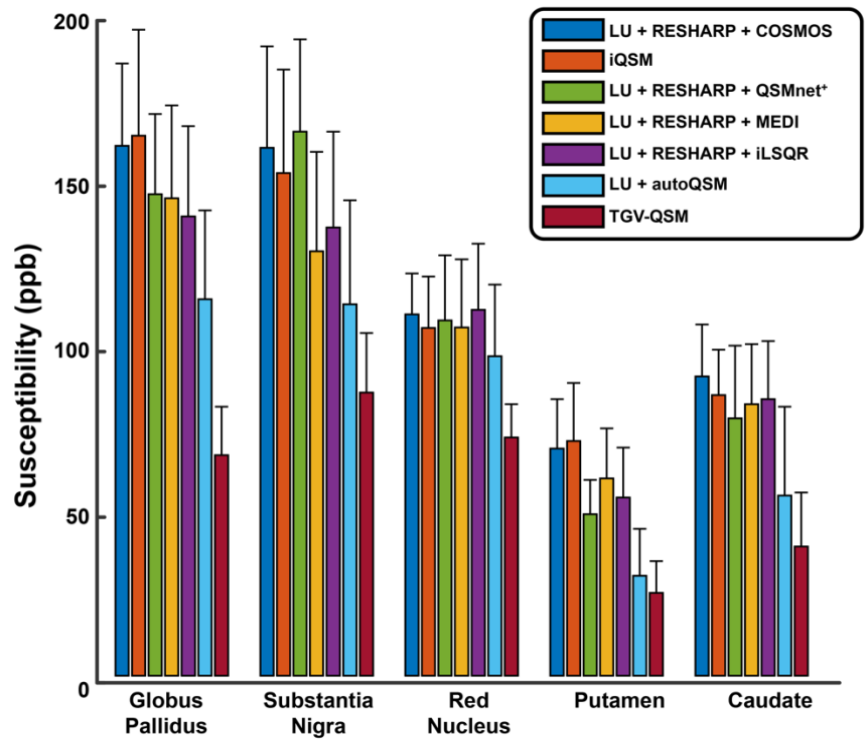

**Figure 7**. ROI analysis of different QSM methods on five deep grey matter regions from four healthy subjects (*in vivo*). The iLSQR, MEDI, and QSMnet[+] were based on Laplacian unwrapping, followed by RESHARP background removal results, while autoQSM takes the Laplacian unwrapping results as its input. Whiskers on the bars indicate the standard deviation of the susceptibility measurements across all subjects. LU denotes Laplacian unwrapping.

COSMOS maps from one subject acquired with five head orientations at 3T were reconstructed using different local field mapping methods and are compared in Fig. 8. Different background field removal methods were tested on brain masks with 1-voxel and 3-voxel brain edge erosions after BET mask extraction. Substantial artifacts were observed in RESHARP, LBV, and PDF-processed COSMOS results with 1-voxel brain edge erosion, while these artifacts were suppressed after eroding more voxels. On the other hand, the deep learning-based methods, SHARQnet and iQFM, are free of visible artifacts and their performances were relatively independent of brain edge erosions. However, noticeable susceptibility contrast suppression was observed in SHARQnet when compared with the proposed iQFM.

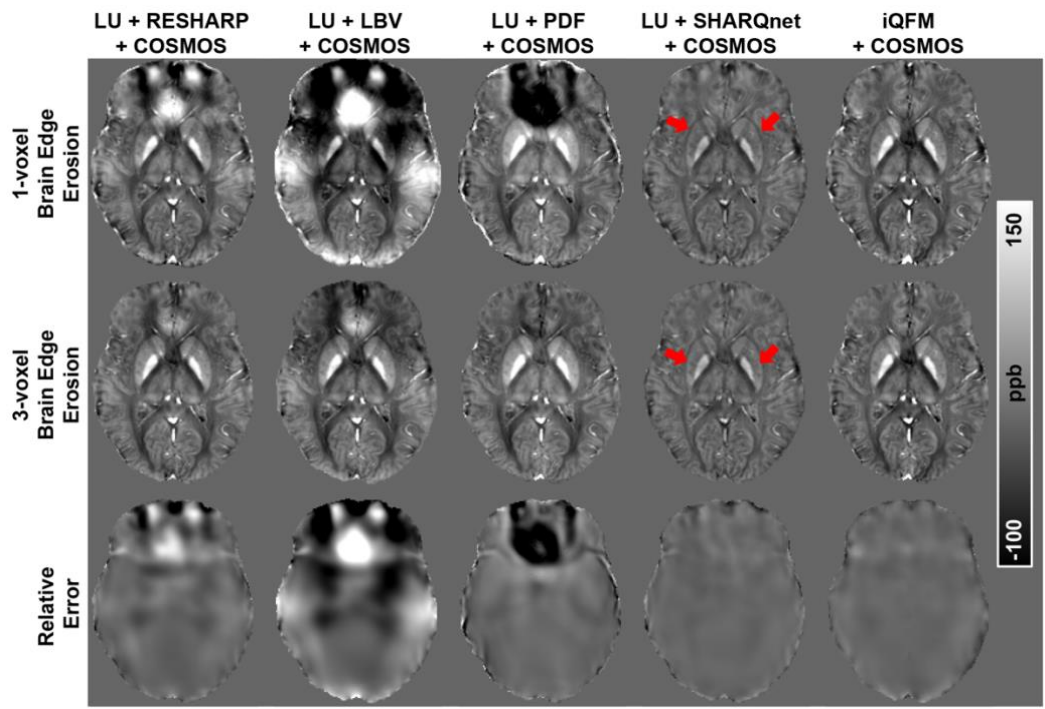

**Figure 8**. Comparison of susceptibility maps computed using the COSMOS approach on results reconstructed from different local field mapping techniques on one healthy subject. The first row illustrates the 1-voxel brain edge erosion results, while the mid-row demonstrates the 3-voxel brain erosion based reconstructions, with relative differences between the two brain

edge erosions shown in the bottom row. Red arrows point to susceptibility underestimation in SHARQnet-based COSMOS results. LU denotes Laplacian unwrapping.

### 3.3. *In vivo* experiment results from ICH and MS patients

QSM results from three ICH patients using different methods are compared in Fig. 9. Traditional iLSQR, MEDI, and TGV-QSM methods led to strong streaking artifacts around hemorrhage sources, as indicated by the red arrows; by contrast, deep learning-based QSMnet$^+$, autoQSM, and iQSM methods resulted in fewer artifacts. Among all methods, the proposed iQSM method produced the most visually appealing susceptibility maps with clear delineations of hemorrhages and minimal streaking artifacts.

Susceptibilities of three hemorrhage regions (yellow arrows in Fig. 9) and one line profile through two adjacent hemorrhages (green line) were measured and plotted in Fig. 10. Average hemorrhage susceptibilities from iQSM (977± 170 ppb) are substantially higher than other methods, including TGV-QSM (710 ± 634 ppb), iLSQR (420 ± 140 ppb), MEDI (520 ± 310 ppb), QSMnet$^+$ (415 ± 130 ppb), and autoQSM (275 ± 88 ppb), which is consistent with simulation results in Fig. 5. In addition, the line profiles also confirmed that iQSM showed the most stable and highest hemorrhage susceptibility measures, while iLSQR, QSMnet$^+$, and autoQSM achieved lower susceptibilities. The MEDI and TGV-QSM results showed the greatest oscillations inside the hemorrhage due to severe artifacts.

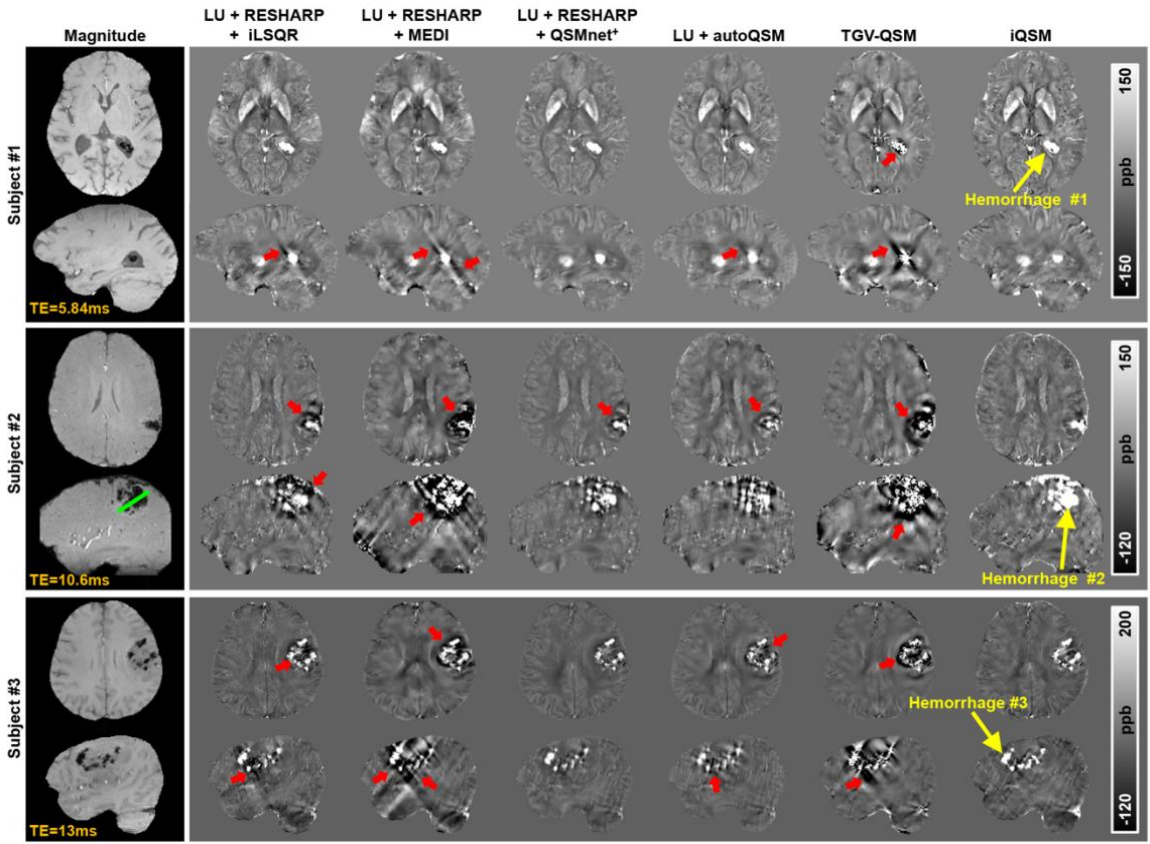

**Figure 9**. Comparison of different QSM methods on three ICH patients. Susceptibility images of two orthogonal views are illustrated for each subject. Red arrows point to the artifacts near the hemorrhage sources in different QSM reconstructions. LU denotes Laplacian unwrapping.

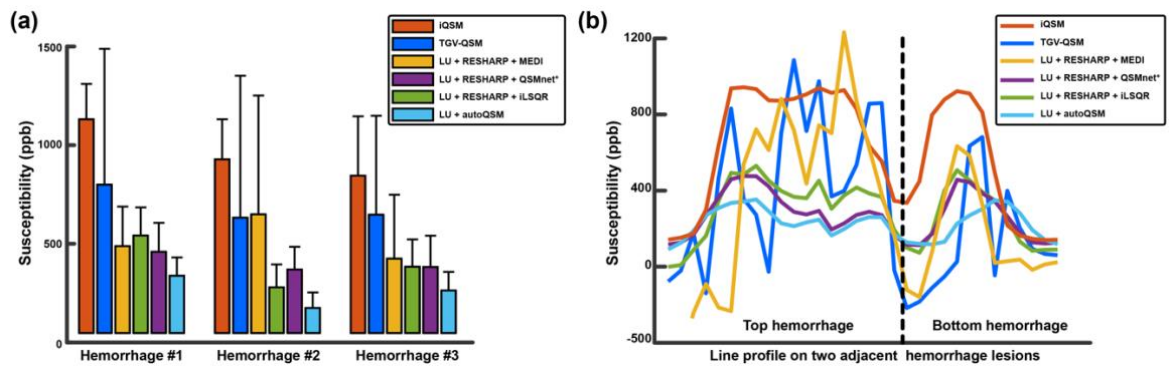

**Figure 10**. Susceptibility measurements of three hemorrhage lesions (located by the yellow arrows in Fig. 9) and line profiles of two adjacent hemorrhage lesions (green line in subject #2 in Fig. 9) from different QSM methods. The iLSQR, MEDI, and QSMnet[+] were based on Laplacian unwrapping, followed by RESHARP background removal results, while autoQSM takes the Laplacian unwrapping results as inputs.

To test the generalization of the deep learning-based methods, QSM reconstructions were performed on two MS patients at 3T with isotropic 1 mm³ resolution. As compared in Fig. 11, three MS lesions were identified in GRE magnitude, FLAIR, and QSM images from all methods, indicated by the red arrows. Another two hyperintensity regions on FLAIR, pointed by the yellow arrows, did not show susceptibility contrast on any QSM results. The TGV-QSM method showed over-smoothed and reduced susceptibility contrast, while the proposed iQSM and MEDI showed the highest contrasts between grey matter, white matter, cerebrospinal fluid, and MS lesions.

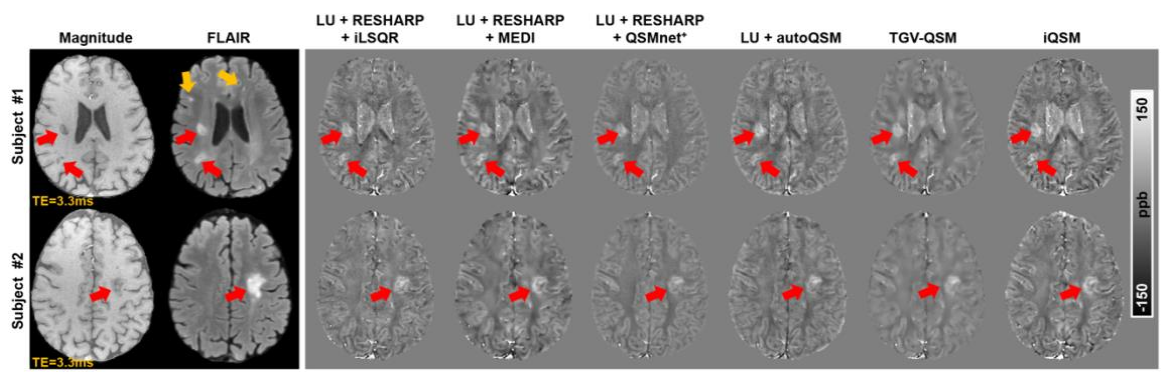

**Figure 11**. QSM results from two MS patients using different reconstruction methods. Red arrows point to the MS lesions identified by visual inspection, while yellow arrows point to the MS lesions not detected by QSM reconstructions. LU denotes Laplacian unwrapping.

## 4. Discussion

This work demonstrated a deep learning method to reconstruct magnetic tissue field and susceptibility directly from the raw phase images for human brain datasets. Specifically, we developed two deep neural networks, iQFM and iQSM, by proposing a novel LoT-Unet architecture for end-to-end near-instant (~130 ms) tissue field and susceptibility mapping. The new networks were compared with established single-step and multi-step methods, including conventional iterative and recent deep learning methods. Results on simulated and *in vivo* datasets from healthy and pathological brains showed that the proposed iQFM and iQSM led to more accurate tissue field and susceptibility maps with the best PSNR, SSIM, NRMSE, and ROI measurements (i.e., deep grey matter, hemorrhage, and calcification). In addition, more robust reconstruction results with substantially reduced streaking artifacts from the high-intensity hemorrhage susceptibility sources were achieved. Despite these improved results on human brains, the feasibility and performances of iQFM and iQSM on MRI phase data from other body regions have not been investigated in this work.

To investigate the benefit of the novel LoT layer, another two neural networks without the LoT layer (i.e., Unet only) were trained for local field and susceptibility mapping using the same datasets. The Unet-only results are compared with iQFM and iQSM in Supp. Fig. 3. Significant artifacts near the phase wraps were observed in Unet-only local field and susceptibility maps (indicated by the red arrows in columns 2 and 3). These artifacts were removed in iQFM and iQSM results (columns 5 and 6). The significant improvement is probably because the LoT layer outputs feature maps free of abrupt discontinuities (i.e., phase wraps), as shown in Supp. Fig. 3 column 4, which dramatically improves the performance of the following Unet. A recent paper (Kames et al., 2021) has successfully recovered the local field map from the homodyne-filtered SWI phase, which is conceptually similar to our iQFM method. However, unlike our method, a 2D network was used to restore the local field slice-by-slice, limiting its feasibility of directly reconstructing susceptibility maps.

Unlike the autoQSM pipeline performing the Laplacian-based unwrapping as a pre-processing step, the iQFM/iQSM method integrates the LoT operations as a convolution-based layer in the network. As a result, the proposed implementation eliminates the errors introduced by explicit Laplacian phase unwrapping (Li et al., 2014; Robinson et al., 2017). To demonstrate this, we have trained another network with the Laplacian-unwrapped phase as the Unet input and compared it with our proposed LoT as the Unet input. As shown in Supp. Fig. 4, Unets taking Laplacian unwrapped phase as inputs resulted in around 10% more error on hemorrhage susceptibility and displayed blooming artifacts near the sinus, compared with our iQFM/iQSM. In addition, the proposed LoT convolutional layer can be efficiently performed using GPUs and integrated with the Unet to form single-step processing. Furthermore, the 27-stencil discrete Laplacian stencil approximation can be replaced with multiple convolutional kernels to be learned from the training data. Specifically, we have trained a new set of iQFM and iQSM networks with a data-driven LoT layer by setting 16 learnable kernels ($\mathbb{R}^{16 \times 1 \times 3 \times 3 \times 3}$) as $K$ in Eq. (3). As a result, 16 feature maps were extracted from a single-channel input (i.e., $sin(\varphi_w)$ or $cos(\varphi_w)$) after the $K$ convolutional operation. Following Eq. (3), $sin(\varphi_w)$ or $cos(\varphi_w)$ was then multiplied to each channel of the 16 feature maps through element-wise multiplication. Subtraction and normalization (i.e., $B_0 \cdot TE$) operations were carried out as the last step of the LoT layer. Examples of the 16 feature maps produced from the full LoT layer (Eq. (3)) for iQFM and iQSM were shown in Supp. Fig. 5. The comparison between the iQFM/iQSM with fixed Laplacian kernels (the third column) and with 16 learnable kernels (the fourth column) is illustrated in Supp. Fig. 4. It is found that the data-driven learned LoT operation further improved the QSM and local field reconstructions. Note that in this paper, all iQFM and iQSM results, if not specified, are based on the iQFM and iQSM networks trained with a standard 27-stencil discrete Laplacian kernel.

Most previous deep learning QSM methods, including QSMnet (Yoon et al., 2018), DeepQSM (Bollmann et al., 2019b), QSMGAN (Chen et al., 2020), and xQSM (Gao et al., 2021), require phase unwrapping and background field removal preprocessing steps. A more recent autoQSM method (Wei et al., 2019) skips the background field removal step but still needs an unwrapped phase as its input. These multi-step QSM pipelines increase reconstruction time and accumulate errors introduced from each step. The iQFM or iQSM methods are extremely fast, requiring around 130 ms for a whole-brain reconstruction (matrix size: 256×256×128), while the multi-step deep learning methods can take from 20 s to more than 4 min.

The proposed iQSM method is the first deep learning-based end-to-end neural network to produce susceptibility maps from the raw phase in one run. This method eliminates the phase unwrapping process, which can be unreliable and underestimating when the phase wraps are severe in high susceptibility regions, such as hemorrhages. As seen in both simulation and *in vivo* experiments, the iQSM method dramatically suppressed the streaking artifacts from hemorrhages, compared with the iteration-based single-step TGV-QSM method.

The proposed iQFM is also the first method to produce a quantitative local field map from the raw phase, which can be used for COSMOS and true-SWI reconstructions and has applications in neurological diseases, such as MS (Elkady et al., 2018; Elkady et al., 2019; Wisnieff et al., 2015). Furthermore, compared with other background field removal methods, iQFM is more robust against brain edge errors and generates similar tissue field maps regardless of different sizes of brain masks. Therefore, the iQFM and iQSM methods can retain more brain edge regions, critical for cortical QSM applications. Indeed, the proposed iQFM and iQSM can be directly computed on the raw phase without any brain tissue extraction step. As shown in Fig. 12, the iQFM and iQSM of the entire FOV can still produce accurate tissue field and magnetic susceptibility in the brain region, compared with results from added brain masking. The robustness of iQFM and iQSM methods against brain masking originates from the convolution scheme of neural networks, also known as the local receptive field design. For each convolutional layer, a small localized local receptive field input is convoluted with the kernel, and thus the brain masking operation only alters a couple of brain edge voxels, as illustrated in Fig. 12 difference maps.

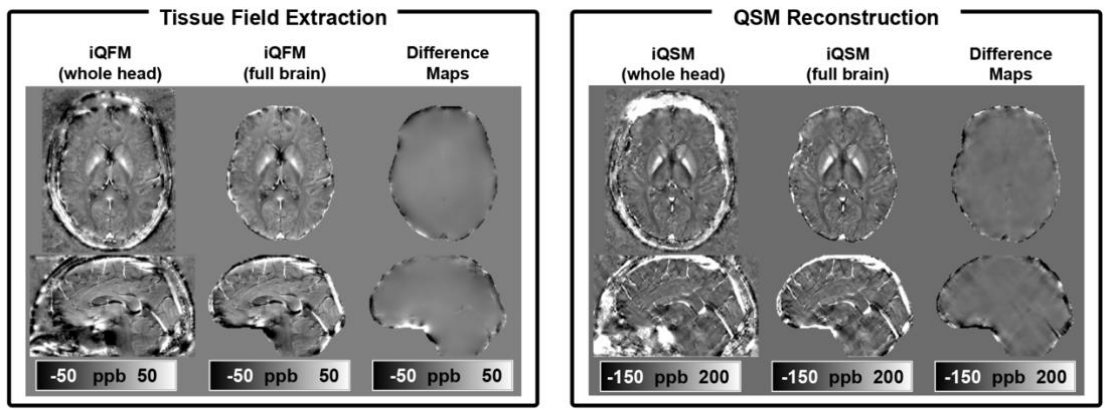

**Figure 12.** Comparison of tissue field and magnetic susceptibility results from iQFM and iQSM on phase images with and without brain tissue extraction. For iQFM and iQSM full brain results, no brain edge erosion was performed after brain extraction using BET. Difference maps of reconstruction results with and without brain masking are shown in the third column.

Most previous deep learning QSM methods chose to train deep neural networks based on the QSM image reconstructed using established conventional methods from the available phase data. With this strategy, the trained networks tend to be more robust against measurement noise. On the other hand, this strategy shares the same drawbacks as the conventional reconstruction algorithms. For example, QSMnet loses the susceptibility anisotropy in white matter regions, and autoQSM leads to underestimated hemorrhage susceptibilities. In this work, we proposed a data generation pipeline to simulate wrapped phase images (as network inputs) from the QSM patches (as network labels), as shown in Fig. 2. The advantages of our forward simulation model compared with directly using the acquired raw phase as the network training input can be summarised as follows. First, it is challenging to acquire data from a sufficient number of hemorrhage patients for network training, while we can generate an abundant and balanced amount of hemorrhage and calcification samples of arbitrary shapes and values for training using our forward simulation model. Second, traditional reconstruction pipelines (e.g., Laplacian unwrapping + RESHARP + iLSQR) tend to underestimate high-susceptibility sources (i.e., hemorrhages), as demonstrated in Fig. 4. Using these underestimated susceptibility results as training labels may bias the neural network and lead to undesirable results. Our simulated training dataset with added pathological susceptibility sources significantly boosted performance on QSM hemorrhages. To demonstrate this, we compared with another set of iQFM/iQSM networks trained on healthy patches only (named iQFM$_{healthy}$ and iQSM$_{healthy}$). It is found that the use of pathological patches can lead to substantial improvement for iQFM/iQSM on high-susceptibility hemorrhages compared to healthy patches only, as shown in Supp. Fig. 6.

Unlike in some previous work (Lai et al., 2020; Wei et al., 2019; Yoon et al., 2018), where testing and training datasets were split from a collection of scans acquired with the same sequence parameters, our testing datasets were obtained from multiple MR centres with different acquisition parameters (i.e., number of echoes, TEs, TRs, flip angles, and field strengths) to ensure a fair comparison between different local field and QSM reconstruction methods. It is found that although the proposed iQFM and iQSM were trained with simulated datasets, they generalized well to *in vivo* acquisitions from both healthy and hemorrhage subjects.

We have also conducted a simulation experiment to compare the robustness of the proposed iQSM and the traditional reconstruction pipeline (i.e., Laplacian unwrapping + RESHARP + iLSQR) against different noise levels, with results shown in Supp. Fig. 7. The simulated noise is added into the complex MRI data:

$$\begin{aligned} M_{noisy} &= |Me^{j\varphi} + n_r + jn_i| \\ \varphi_{noisy} &= \angle(Me^{j\varphi} + n_r + jn_i) \end{aligned} \tag{5}$$

where $M$ and $\varphi$ are the simulated magnitude and wrapped phase without noise; $n_r$ and $n_i$ are two independent Gaussian-distributed noise added into the real and imaginary parts of the complex MRI data to simulate the noise-contaminated magnitude $M_{noisy}$ and wrapped phase $\varphi_{noisy}$. Four different SNR levels (i.e., 80, 50, 25, 10) were tested for comparison. As shown in Supp. Fig. 7, the performance of the conventional QSM pipeline degrades substantially with decreased SNRs, while the iQSM method resulted in a more robust susceptibility mapping of both hemorrhage and normal brain tissue with suppressed noise presented in the final QSM results.

In this work, all performance evaluations were conducted on experiments of isotropic 1 mm$^3$ resolution, which is also the image resolution of the training datasets. For applications of image resolutions different from training datasets, image interpolation (resample these data into 1 mm isotropic) is necessary. Otherwise, the proposed iQFM and iQSM will need retraining with datasets of matched image resolution, and the Laplacian kernel in the LoT layer also needs to be modified with a finite-difference approximation of second-order derivatives with non-uniform spacing for anisotropic voxel sizes. Note that iQFM and iQSM with learnable kernels in LoT layer can automatically learn the values of the LoT kernels from the training data and do not need this explicit modification for the convolutional kernels. In addition, even though the iQFM is independent of acquisition orientations, the iQSM needs image rotation, interpolation, and re-slicing for phase data acquired at oblique orientations, which may blur the images. Future work should investigate deep learning strategies for reconstructing QSM of arbitrary image resolution and acquisition obliquity to the main magnetic field.

## 5. CONCLUSION

This study proposed a novel deep learning-based end-to-end framework, LoT-Unet, directly reconstructing tissue field and susceptibility maps from the raw wrapped phase through two neural networks (i.e., iQFM and iQSM). Experimental results showed that iQFM and iQSM led to more accurate and robust tissue field and susceptibility reconstructions, especially for intracranial hemorrhages of high susceptibilities, compared with established single-step and multi-step methods. In addition, the iQFM and iQSM methods can retain the full brain without eroding brain edges, critical for cortical QSM applications. Lastly, the proposed neural networks substantially accelerate the QSM reconstruction time from minutes using multi-step methods to 130 milliseconds.

## DATA AND CODE AVAILABILITY STATEMENTS

Data are available on request due to privacy/ethical restrictions. Source codes and trained networks are available at: https://github.com/sunhongfu/deepMRI/tree/master/iQSM.

## ACKNOWLEDGMENTS

VV acknowledges support from the Australian Research Council (DP190101889 and IC170100035).

GBP acknowledges support from the Canadian Institutes for Health Research (CIHR FDN-143290) and the Campus Alberta Innovates Program (RC-12-003).

HS acknowledges support from the Australian Research Council (DE210101297).

The authors would like to thank Dr Markus Barth for his helpful discussion.